\documentclass[conference]{IEEEtran}

\usepackage{amsmath}
\usepackage{graphicx}
\usepackage{color}
\usepackage{cite}

\begin{document}
	
	
	\graphicspath{{images/}}

	\title{3-D Super-Resolution Ultrasound (SR-US) Imaging \\ using a 2-D Sparse Array \\ with High Volumetric Imaging Rate}
	
	\author{\IEEEauthorblockN{Sevan~Harput$^1$, Kirsten Christensen-Jeffries$^2$, Jemma Brown$^2$, Jiaqi Zhu$^1$, Ge Zhang$^1$, Chee Hau Leow$^1$, \\ Matthieu Toulemonde$^1$, Alessandro Ramalli$^{3,4}$, Enrico Boni$^3$, Piero Tortoli$^3$, \\ Robert J. Eckersley$^{2*}$, Chris Dunsby$^{5*}$, and Meng-Xing~Tang$^{1*}$}
		\IEEEauthorblockA{ \small \\
			\small $^1$ULIS Group, Department of Bioengineering, Imperial College London, London, SW7 2BP, UK  \\
			\small $^2$Biomedical Engineering Department, Division of Imaging Sciences, King's College London, SE1 7EH, London, UK  \\
			\small $^3$Department of Information Engineering, University of Florence, 50139 Florence, IT  \\
			\small $^4$Lab. on Cardiovascular Imaging \& Dynamics, Dept. of Cardiovascular Sciences, KU Leuven, Leuven, Belgium \\
			\small $^5$Department of Physics and the Centre for Pathology, Imperial College London, London, SW7 2AZ, UK  \\
			\small $^*$These authors contributed equally to this work  \\
			E-mail: S.Harput@imperial.ac.uk, Mengxing.Tang@imperial.ac.uk}	
	}
	
	\maketitle

	\begin{abstract}
		\boldmath
		Super-resolution ultrasound imaging has been so far achieved in 3-D by mechanically scanning a volume with a linear probe, by co-aligning multiple linear probes, by using multiplexed 3-D clinical ultrasound systems, or by using 3-D ultrasound research systems. In this study, a 2-D sparse array was designed with 512 elements according to a density-tapered 2-D spiral layout and optimized to reduce the sidelobes of the transmitted beam profile. High frame rate volumetric imaging with compounded plane waves was performed using two synchronized ULA-OP256 systems. Localization-based 3-D super-resolution images of two touching sub-wavelength tubes were generated from a 120 second acquisition.

	\end{abstract}
	
	\maketitle

	\section{Introduction}
	
By localizing spatially isolated microbubbles through multiple frames, super-resolution ultrasound (SR-US)  images can be generated. Even at ultrasonic frequencies in the low MHz range, a localization precision of a few micrometers can be achieved~\cite{Viessmann2013,Desailly2015}. In the absence of sample motion, it is this localization precision that determines the maximum achievable resolution in super-resolution images. If motion is present and subsequently corrected post-acquisition, then the motion correction accuracy can also limit the achievable spatial resolution~\cite{Harput2017a,Harput2018}. Researchers demonstrated the use of 2-D SR-US imaging in many studies using microbubbles~\cite{Christensen-Jeffries2015,Ackermann2016,Bar-Zion2017,Foiret2017,Harput2017b,Couture2018,Song2018,Opacic2018,Ilovitsh2018} and nanodroplets~\cite{Luke2016,Yoon2018,Zhang2018}. All of these studies were based on imaging with 1-D probes that can only display 2-D slices of a 3-D structure, thus making the volumetric observations more challenging. In addition to this, 2-D ultrasound imaging introduces more limitations in localization based super-resolution imaging. First, super-resolution cannot be achieved in the elevational direction. Second, out-of-plane motion can only be compensated for movements smaller than the elevational beamwidth of the transducer. However, with the implementation of 3-D super-resolution ultrasound imaging, diffraction limited resolution can be overcome in every direction and 3-D motion can be tracked over larger scales.

Although 3-D super-resolution imaging has not been achieved with a 2-D imaging probe with a high volumetric rate, several studies have extended super-resolution into the third dimension. O'Reilly and Hynynen used a subset of a hemispherical transcranial therapy array to generate 3-D super-resolution images of a spiral tube phantom through an \textit{ex vivo} human skullcap~\cite{Reilly2013}. Desailly \textit{et al.} used 2 parallel series of transducers to image microfluidic channels and obtained 3-D super-localization by fitting parallel parabolas in the elevation direction~\cite{Desailly2013}. Errico \textit{et al.} performed a coronal scan of an entire rat brain and generated 2-D super-resolution slices by using a micro-step motor and a linear array~\cite{Errico2015}. Lin \textit{et al.}  mechanically scanned a rat FSA tumor using a linear array mounted on a motorized precision motion stage to generate 3-D super-resolution images by calculating the maximum intensity projection from all 2-D slices~\cite{Lin2017}. Christensen-Jeffries \textit{et al.} generated volumetric 3-D super-resolution images of cellulose tubes at the overlapping imaging region of two orthogonal linear arrays at the focus~\cite{Christensen-Jeffries2017a}.

	\begin{figure*}[tb]
		\centering
		\includegraphics[viewport = 80 80 1100 350,  width = 160mm, clip]{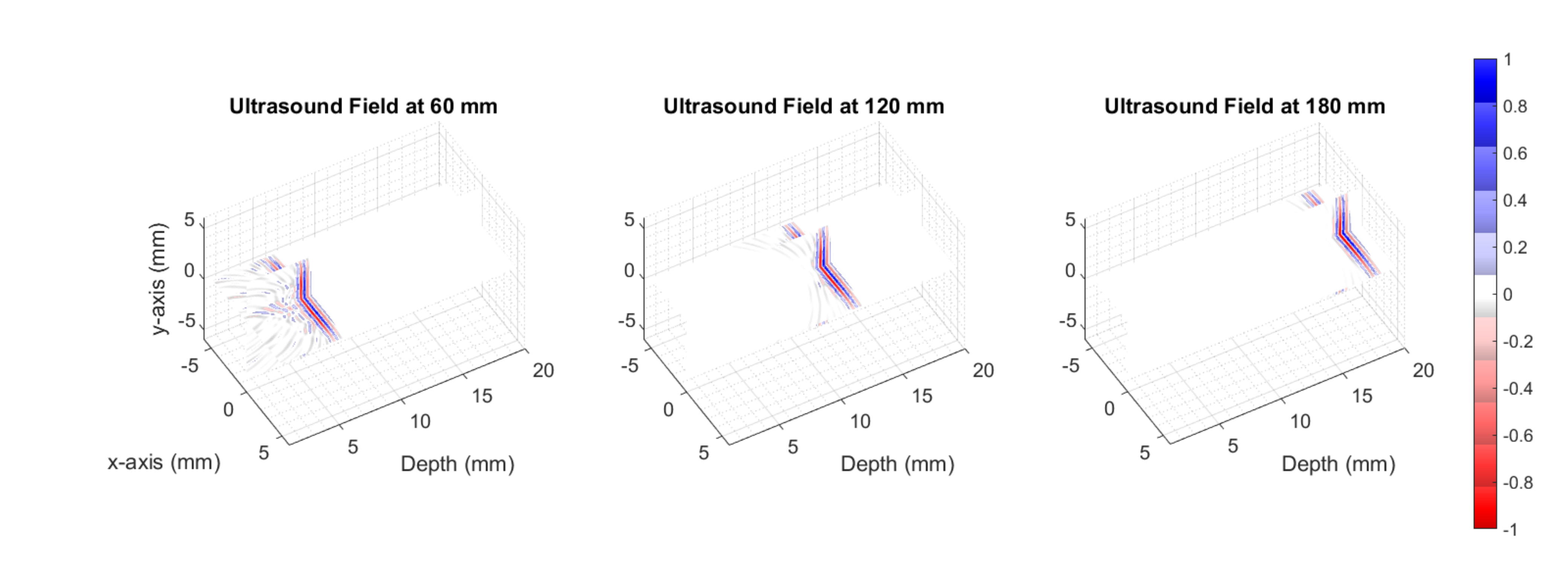}
		\caption{Simulated ultrasound field propagation from the sparse array computed at 6, 12 and 18~mm depths.}
		\label{fig:SparseArray_Tukey_Plane_wave_propogation}
	\end{figure*}

Using a full 2-D array for 3-D SR-US imaging requires a system with very large number of independent channels, the design of which might be impractical due to the high cost, complexity, and large data size. Several technologies have been developed to use a large number of active elements with fewer scanner channels in order to simplify ultrasound systems and probes. In this study, a density-tapered sparse array method was chosen instead of a full 2-D array to reduce the number of channels and hence the amount of data while maintaining the frame rate. This approach is similar to previous studies on minimally redundant 2-D arrays~\cite{Karaman2009} and sparse 2-D arrays~\cite{Austeng2002,Diarra2013,Roux2016,Roux2017,Roux2018}, but uses more number of elements to improve transmit power and receive sensitivity. Our method significantly differs from row-column addressing and  multiplexing approaches since it maintains simultaneous access to all probe elements through independent channels. The sparse array was designed specifically for high volumetric rate 3-D super-resolution ultrasound imaging based on a density-tapered spiral layout~\cite{Ramalli2015a}.

	\section{Materials and Methods}
	
	\subsection{2-D Sparse Array}
	
	\begin{figure}[!t]
		\centering
		\includegraphics[viewport = 30 35 540 549,  width = 54mm, clip]{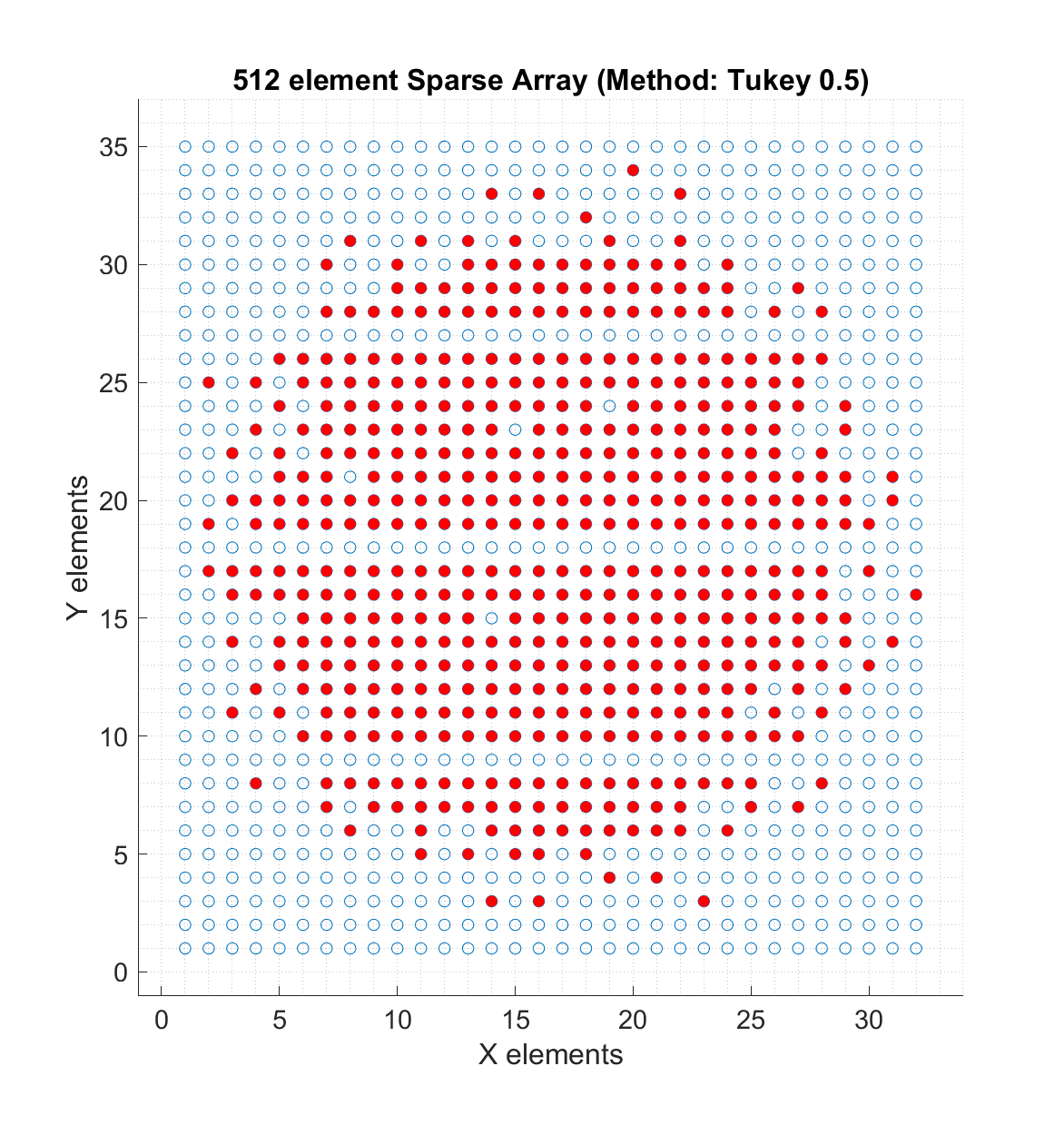}
		\caption{Layout of the 2-D sparse array. Empty rows are introduced due to manufacturing limitations and not related to the density-tapered 2-D spiral method.}
		\label{fig:SparseArray}
	\end{figure}

A sparse array was designed by selecting 512 elements from a $35 \times 32$ element 2-D matrix array as shown in Fig.~\ref{fig:SparseArray}. Row number 9, 18, and 27 are intentionally left blank for wiring. The method to select the location of sparse array elements is based on the density-tapered 2-D spiral layout. This method arranges the elements according to Fermat's spiral seeds with spatial density modulation and reduces the sidelobes of the transmitted beam profile. This deterministic, aperiodic, and balanced positioning procedure guarantees uniform performance over a wide range of imaging angles~\cite{Ramalli2015a}.
	
To verify the feasibility of the sparse array for 3-D imaging with plane waves, Field II simulations were performed~\cite{Jensen1992,Jensen1996}. [29].	Ultrasound propagation from the sparse array is simulated at different depths as shown in Fig.~\ref{fig:SparseArray_Tukey_Plane_wave_propogation}. Closer to the transducer the radiated plane wave has a tail, which makes it not suitable for plane wave imaging. However, far from the transducer, the ultrasound field becomes uniform and the residual field behind the wavefront is diminished, which is suitable for plane wave imaging.
	
After the \textit{in silico} feasibility test, the 2-D sparse array (Vermon S.A., Tours, France) was specifically fabricated for 3-D SR-US with an element size of $300 \times 300$ $\mu$m, center frequency of 3.7 MHz and a bandwidth of 60\%.

	\subsection{Experimental Setup}
	
Two ULA-OP256~\cite{Boni2016,Boni2017} systems were synchronized to transmit 9 plane waves steered within a range of $\pm10$ degrees in the lateral and elevational directions from the 512 selected elements. Each compounded volume consisted of 9 volumetric datasets acquired in 3.6 ms with a pulse repetition frequency of 2500 Hz. This compounded plane wave imaging scheme was used to measure the super-localization precision of the system before the experiments, which was found to be 18~$\mu$m.

\begin{figure}[!t]
		\centering
		\includegraphics[viewport = 40 470 390 650,  width = 70mm, clip]{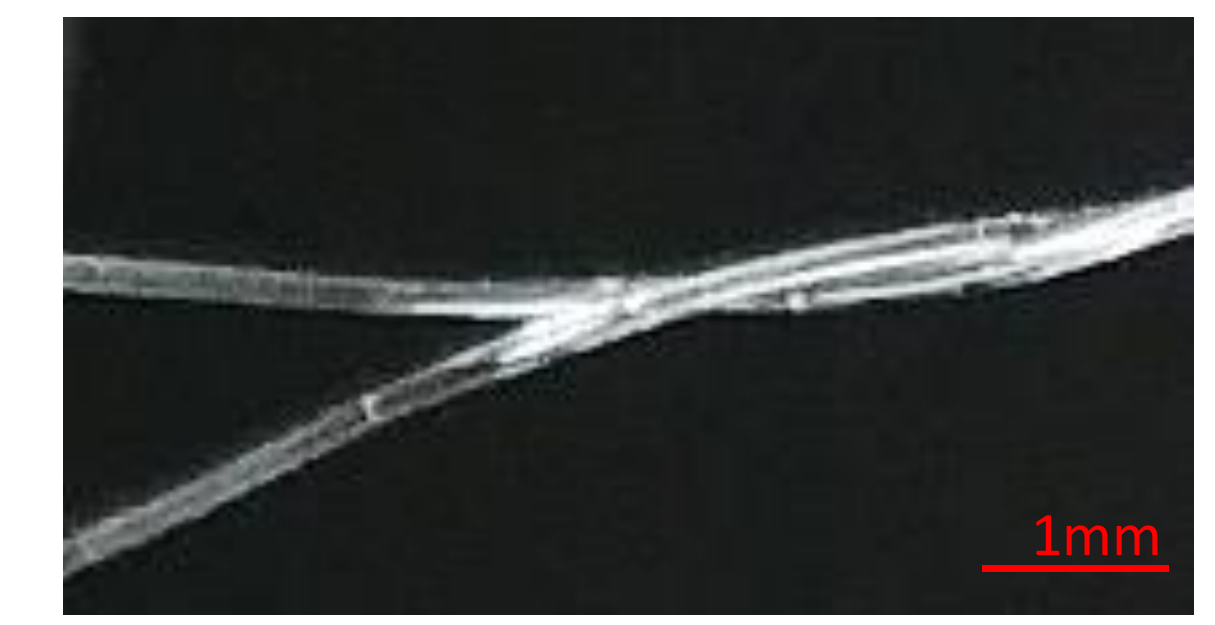}
		\caption{Optical Images of two 200 $\mu$m cellulose tubes arranged in a double helix pattern.}
		\label{fig:Optical_image}
	\end{figure}

Two 200 $\mu$m tubes arranged in a double helix shape were imaged during the experiments as shown in Fig.~\ref{fig:Optical_image}. A 1:1000 diluted Sonovue (Bracco S.p.A, Milan, Italy) solution was flowed through both tubes with a constant flow rate that produced an average microbubble velocity of 10~mm/s. A total of 3000 volumetric ultrasound frames were acquired in 120 seconds. Beamforming and volumetric reconstruction were performed offline. SVD was used to separate the microbubble signals from the echoes originating from the tube and the assembly holding the tube. Localization of isolated microbubbles was performed on every acquired volume to generate the 3-D super-resolved volumes~\cite{Christensen-Jeffries2017}.

Fast data acquisition was used with a PRF of 2500 Hz to ensure minimum artefacts on compounded volumes due to moving microbubbles. Before the next fast acquisition, there was a 40~ms interval to allow the microbubbles to travel approximately half the size of the B-mode PSF. This short pause between acquisitions was introduced to reduce the data redundancy and maximize the total acquisition duration for the allocated memory size.

	\section{Results \& Discussion}
	
\begin{figure}[!t]
	\centering
	\includegraphics[viewport = 0 40 690 434,  width = 76mm, clip]{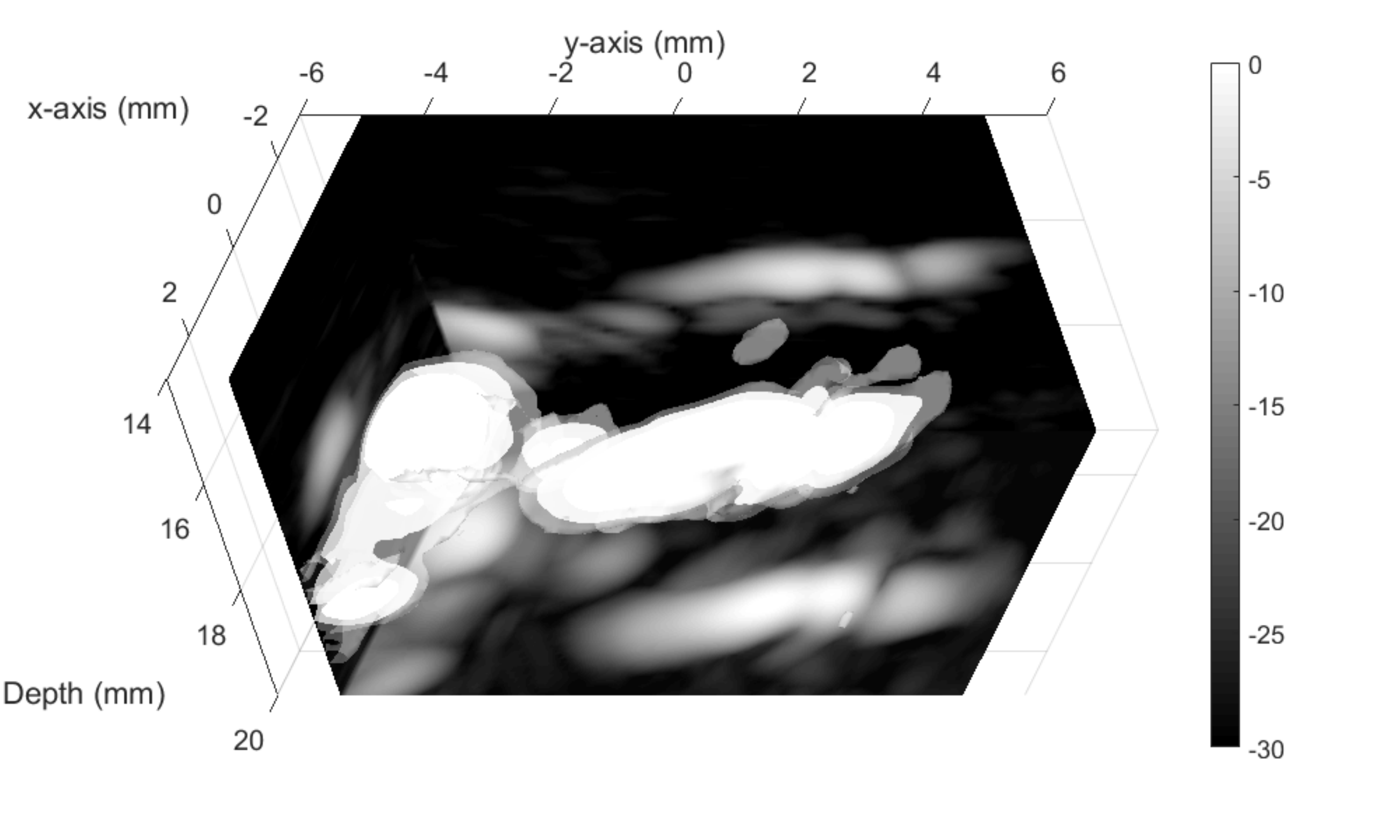}
	\caption{Isosurface of the 3-D ultrasound image is plotted in white at -15, -20, and -25 dB levels with degrading transparency. 2-D maximum intensity projections with a 30~dB dynamic range is overlaid on the volumetric image.}
	\label{fig:Bmode_3D}
\end{figure}

Fig.~\ref{fig:Bmode_3D} shows the 3-D visualization of microvessels overlaid on 2-D maximum-intensity-projection (MIP) slices, where it is not possible to visualize two 200 $\mu$m tubes. The full-width-half-maximum (FWHM) of the 3-D B-mode point-spread-function (PSF) was measured by using linear interpolation as 793, 772, and 499 $\mu$m in X, Y, Z directions respectively~\cite{Harput2014a}.

\begin{figure*}[!t]
	\centering
	\includegraphics[width = 180mm]{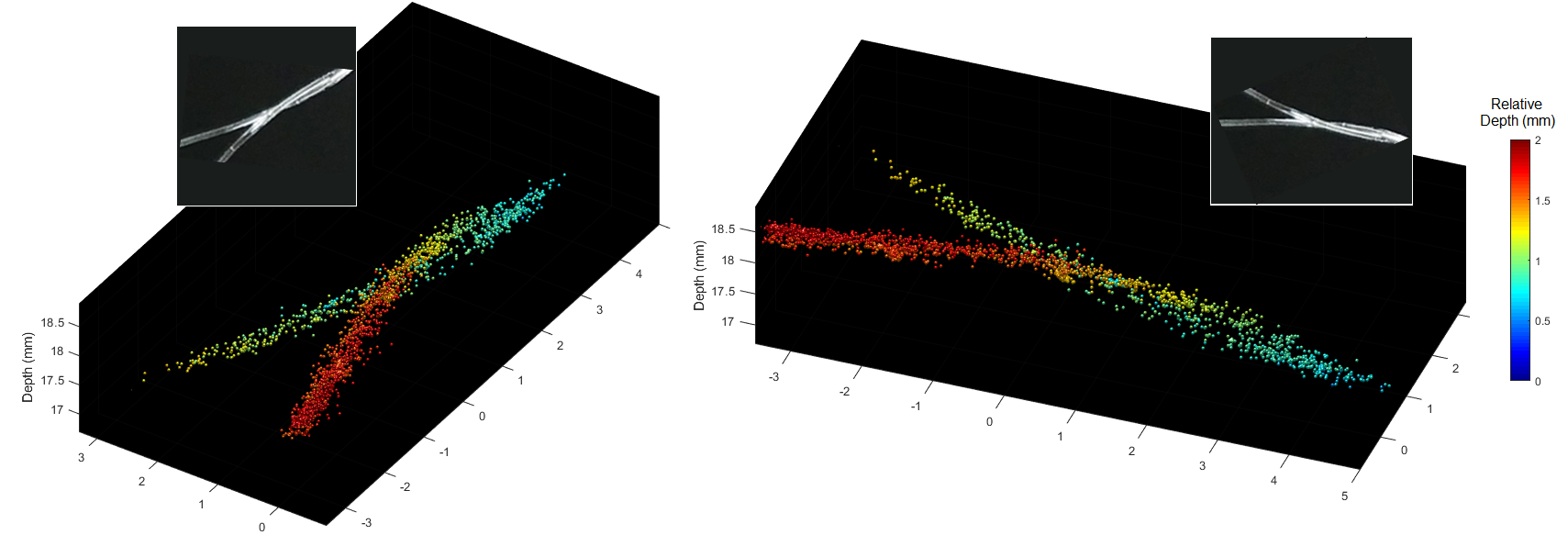}
	\caption{Two images zoom show the structure from different viewing angles. Depth-encoded colorscale is added to improve the visualization of two tubes arranged in a double helix shape. The corresponding optical images are presented as an inset.}
	\label{fig:SR_3D_colorcoded}
\end{figure*}

Fig.~\ref{fig:SR_3D_colorcoded} shows the 3-D SR-US images of the sub-wavelength structures, where the imaging wavelength is 404 $\mu$m in water at 25$^{\circ}$C. A total of 2319 microbubbles were localized within the 3000 volumes after compounding. Due to the large number of localizations, the 3-D structure of the tubes cannot be clearly visualized in a single 2-D image. To improve the visualization, 3-D SR-US images are plotted with depth information color-coded in the image from different viewing angles. The FWHM of a single tube, which was measured over a 2~mm length, appeared as 193 $\mu$m at the widest point and 81\% of the super-localizations were within a diameter of 200 $\mu$m.

Volumetric imaging with 2-D sparse arrays can be a reasonable choice instead of using a full matrix array or a multiplexed array. In comparison to full matrix arrays, the proposed solution is less expensive; it can achieve the same volumetric acquisition speed since all elements of 2-D spiral array are continuously connected to every channel in the system; and it has a comparable spatial resolution since the bandwidths and aperture sizes are almost the same. In comparison to 2-D multiplexed arrays, it has the flexibility to modify the number of compoundings depending on the application. When the acquired ultrasound volume data is a combination of multiple transmissions and acquisitions, intra-volume motion artefacts will be generated. This can be a problem for multiplexed 3-D ultrasound systems for some applications.

	\section{Conclusion}
	
The main limitation of localization-based SR-US imaging performed in 2-D is the lack of super-resolution in the elevation direction. In this study, this issue was addressed by using a bespoke 2-D sparse array, which achieved super-resolution in the elevational direction and sufficient SNR for 3-D SR-US imaging. Two 200 $\mu$m, smaller than half wavelength, tubes arranged in a double helix shape were resolved in 3-D using the 2-D spiral array probe.
	
	\section*{Acknowledgments}  
	   
This work was supported mainly by the EPSRC under Grant EP/N015487/1 and EP/N014855/1, in part by the King's College London (KCL) and Imperial College London EPSRC Centre for Doctoral Training in Medical Imaging (EP/L015226/1), in part by the Wellcome EPSRC Centre for Medical Engineering at KCL (WT 203148/Z/16/Z), in part by the Department of Health through the National Institute for Health Research comprehensive Biomedical Research Center Award to Guy's and St Thomas' NHS Foundation Trust in partnership with KCL and King's College Hospital NHS Foundation Trust, in part by the Graham-Dixon Foundation and in part by NVIDIA GPU grant.

	
	\bibliography{BuBBle,Ultrasound,SignalProcessing,MotionCorrection,SuperRes,3D_Imaging}       
	\bibliographystyle{IEEEtran}

\end{document}